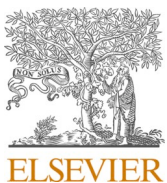
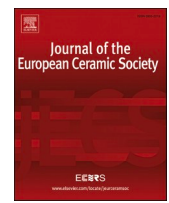

# Effect of organic solvent on the cold sintering processing of $SrFe_{12}O_{19}$ platelet-based permanent magnets

Aida Serrano [a,*], Eduardo García-Martín [a,b], Cecilia Granados-Miralles [a], Jesús López-Sánchez [c], Giulio Gorni [d], Adrián Quesada [a], José F. Fernández [a]

[a] *Departamento de Electrocerámica, Instituto de Cerámica y Vidrio (ICV), CSIC, 28049 Madrid, Spain*
[b] *Departamento de Física de Materiales, Universidad Complutense de Madrid, 28040 Madrid, Spain*
[c] *Instituto de Magnetismo Aplicado, UCM-ADIF, 28230 Las Rozas, Spain*
[d] *CELLS-ALBA Synchrotron Light Facility, 08290 Cerdanyola del Vallès, Spain*

**A B S T R A C T**

In this work we have investigated the effect of the solvent during the processing of $SrFe_{12}O_{19}$ platelet-based permanent magnets by cold sintering process (CSP) plus a post-thermal treatment. Several organic solvents: glacial acetic acid, oleic acid and oleylamine have been analyzed, optimizing the CSP temperatures at 190−270 °C, under pressures of 375−670 MPa and 6−50 wt% of solvent. Modifications in the morphological and structural properties are identified depending on the solvent, which impacts on the magnetic response. Independently of the solvent, the mechanical integrity of ferrite magnets obtained by CSP is improved by a post-annealing at 1100 °C for 2 h, resulting in relative densities around 92 % with an average grain size of 1 μm and a fraction of $SrFe_{12}O_{19}$ phase >91 %. $H_C \geq 2.1$ kOe and $M_S$ of 73 emu/g are obtained in the final sintered ceramic magnets, exhibiting the highest $H_C$ value of 2.8 kOe for the magnet sintered using glacial acetic acid.

## 1. Introduction

The sintering process is the usual methodology employed in the manufacturing of ceramic pieces to densify them from a green body. Among all the available possibilities to sinter a material, the recent cold sintering process (CSP) stands out for its attractive potential [1]. The CSP is based on the mixture of the inorganic compounds in powder form with a liquid phase that can partially solubilize high chemical potential regions of the particles, facilitating their rearrangement, interdiffusion and promoting mass transport under uniaxial pressure is applied at low temperatures [2–7]. The densification process during CSP is given by the dissolution, which can be congruent or incongruent, and the re-precipitation process, where Ostwald ripening and recrystallisation phenomena can be involved [4,5]. The CSP allows varying a large number of processing parameters to optimize the final compounds [2,4], and, in some instances, the process may be assisted by a post-thermal treatment to achieve the densification of materials in a controlled way as well as tailor the final properties of the sintered pieces [2–4,6].

One of the most important parameters during the CSP is the nature and quantity of the transient liquid employed as solvent. A large number of liquids including water, alcohols, organic acids or mixtures with a boiling point below 200 °C and in the range 1−25 wt% have been reported [4]. However, solvents in other percentages and boiling point can also be considered in order to achieve the densification of the specific materials.

Currently, a great number of investigations have reported on the CSP of a large number of binary, ternary and quaternary compounds, allowing a considerable reduction of operating temperatures and times in the sintering processes of several materials, and as a consequence, the reduction in the processing energy (associated $CO_2$ footprint). The CSP is a sustainable manufacturing process that has also been used for the fabrication of several composite structures from dissimilar materials [8–11], and it can be considered to sinter numerous advanced materials. In fact, our group has recently reported for the first time the sintering process of ceramic magnets based on Sr hexaferrites by CSP plus a subsequent post-annealing process [12]. At present, several investigations on CSP-assisted hexaferrites have been addressed. For instance, Lowum et al. have densified $BaFe_{12}O_{19}$ at 300 °C under pressure of 530 MPa by hydroflux-assisted densification technique based on CSP to facilitate particle consolidation [13]. Specifically on cold sintering of $SrFe_{12}O_{19}$, Rajan et al. have studied $Li_2MoO_4$–$SrFe_{12}O_{19}$ composites under a pressure of 450 MPa and a temperature of 200 °C using water as liquid medium for the sintering process [14].

The interest in hexaferrites-based magnets lies in the fact that they present remarkable structural and hard magnetic properties, making them the most widely used permanents magnet materials in the world, amounting to 80 % of the volume of magnets manufactured annually [15–22]. Hexaferrites are extensively employed due to their easy production and corrosion resistance showing a stability of operation over time, temperature, and radiation [23]. In addition, their availability, cost and carbon footprint make them an important alternative to rare-earth permanent magnets, offering an interesting optional source of green energy to be employed in a large number of applications such as wind turbines, microwave devices, small electric motors, recording media, magneto-optics, biomarkers, and biosensors [24–29]. As a consequence, their demand is expected to increase in the next few years. In this framework, the development of new and greener sintering routes of the hexaferrites is of extreme interest.

Herein, we have studied the impact of the solvent employed on the sintering process of ceramic magnets based on $SrFe_{12}O_{19}$ powders in order to optimize their final properties. For that, glacial acetic acid, oleic acid and oleylamine have been considered in the sintering route consisting of a CSP stage plus a subsequent annealing at 1100 °C. The CSP parameters have been adjusted according to the solvent employed in order to achieve the best density values. The properties of the samples have been evaluated after each stage of the sintering process, identifying specific morphological and structural characteristics depending on the solvent employed and the process stage, which remarkably affect the magnetic response.

## 2. Experimental methods

$SrFe_{12}O_{19}$ powders from Max Baermann Holding (Germany) [30], with platelet shape and a bimodal particle size of 100–500 nm and 1–3 μm (see Supporting Information, SI), were employed to sinter the permanent magnets using several organic solvents. For that, a sintering process consisting of CSP plus a subsequent post-annealing step was followed, as previously reported in [11,12]. A representative scheme is shown in Fig. 1A. Firstly, magnetic powders were dispersed by a 10 min-long dry dispersion method [31,32] and afterwards, they were manually mixed with a pure organic solvent for 10 min. The selected organic solvents were: glacial acetic acid, oleic acid and oleylamine. The resulting granulated powders obtained from the mixture were first pressed at 1 bar for 5 min at room temperature in a cylindrical die with an inner diameter of 0.83 mm and consequently submitted to CSP in a BURKLE D-7290 press, heating at specific CSP conditions of temperature, pressure and time ($T_{CSP}$, $P_{CSP}$ and $t_{CSP}$). The annealing rate during the CSP was 20 °C/min and the subsequent cooling down was in air. The CSP parameters and the powder to solvent ratio were optimized depending on the solvent looking for the highest density values of the magnets. Finally, all the CSP pieces were post-annealed at 1100 °C for 2 h in an air atmosphere as final step, with a heating rate of 5 °C/min. Table 1 presents the selected samples obtained after the CSP using glacial acetic acid (GAA), oleic acid (OA) and oleylamine (OL), and after the post-annealed process using glacial acetic acid (GAA1100), oleic acid (OA1100) and oleylamine (OL1100). The specific conditions during the CSP and the solvent percentage are indicated in each case.

All samples were investigated after the CSP step and the post-thermal treatment step at 1100 °C. The relative density of the $SrFe_{12}O_{19}$ ceramics was calculated by the mass/dimensions values and also by the Archimedes method. For each piece, relative density values were obtained considering the compositional phases identified by X-ray diffraction (XRD) and X-ray absorption spectroscopy (XAS) as well as their theoretical density values: 5.10 g/cm³ for $SrFe_{12}O_{19}$ [33], 5.26 g/cm³ for α-$Fe_2O_3$ [34], and 5.01 g/cm³ for SrO [35]. Taking into account that amorphous solids can exhibit up to 15 % lower density than crystalline materials [36], the relative density value calculated for the sample with the highest concentration of amorphous phase would correspond to 3 % higher (GAA sample). For the rest of the samples the error is lower than 1 %.

**Table 1**
Description of samples sintered for this work using different pure organic solvents after the CSP step and the post-annealing step at 1100 °C for 2 h. The solvent, the sintering step and the CSP parameters ($T_{CSP}$, $P_{CSP}$ and $t_{CSP}$) are listed.

| Sample | Pure solvent, % | Sintering process | CSP conditions: $T_{CSP}$, $P_{CSP}$, $t_{CSP}$ |
|---|---|---|---|
| GAA | Glacial acetic acid, 50 % | CSP | 190 °C, 375 MPa, 2 h |
| GAA1100 | | CSP + post-annealing | |
| OA | Oleic acid, 8 % | CSP | 270 °C, 670 MPa, 3 h |
| OA1100 | | CSP + post-annealing | |
| OL | Oleylamine, 6 % | CSP | 260 °C, 670 MPa, 2 h |
| OL1100 | | CSP + post-annealing | |

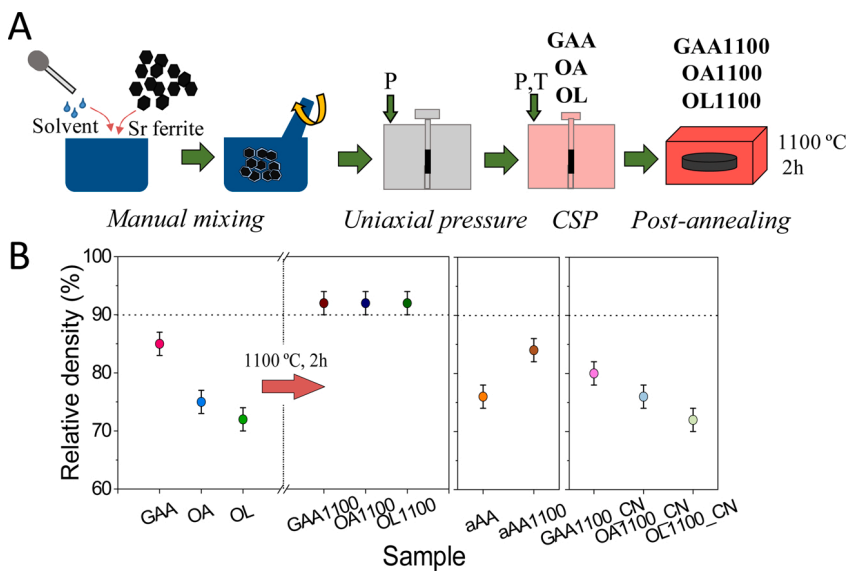

**Fig. 1.** A) Scheme of the sintering process followed for sintering dense $SrFe_{12}O_{19}$ ceramics in this work. B) Relative densities of pieces modifying the organic solvent after the CSP: GAA, AO and OL, and after the post-annealing at 1100 °C for 2 h: GAA1100, OA1100 and OL1100. Results are compared with samples prepared by CSP using an aqueous acetic acid, aAA, and after the post-annealing process, aAA1100, as well as with samples fabricated by conventional route at 1100 °C for 2 h employing the same organic solvents: GAA1100_CN, OA1100_CN and OL1100_CN.

The morphology of samples was evaluated by field emission scanning electron microscopy (FESEM), with an S-4700 Hitachi instrument at 20 kV on fresh fractured surfaces. ImageJ software was used to analyze the average particle size from the particle size distribution obtained from the FESEM images.

XRD experiments were performed in a D8 Advanced Bruker diffractometer using a Lynx Eye detector and a Cu Kα radiation (λ = 0.154 nm) in the 2θ range of 25–65 deg. Rietveld refinements of the XRD data were carried out using the software FullProf [37]. The uncertainties on the refined parameters are calculated considering propagation of errors. In the refinements, a Thompson-Cox-Hastings pseudo-Voigt function was used to model the peak-profile [38]. The instrumental-contribution to the peak-broadening was determined from measurements of a standard powder (NIST LaB6 SRM® 660b) [39] and deconvoluted from the data. The sample-contribution to the broadening was considered as purely size-originated. In all samples, quantitative information on the crystalline compositional phases, the lattice parameters and the average crystallite size was extracted from Rietveld analysis.

Confocal Raman microscopy experiments were performed at room temperature using a Witec ALPHA 300RA confocal Raman microscopy with a linearly p-polarized Nd:YAG laser (532 nm). Raman measurements were performed using an objective lens with a numerical aperture of 0.95 and fixing the laser excitation power at 0.5 mW in order to avoid overheating or damaging in the samples. An average Raman spectrum was attained for each sample from an *in-plane* mapping of $10 \times 10 \ \mu m^2$ on the sample surfaces, where each punctual Raman spectrum is recorded every 200 nm. Several mappings were carried out to evaluate the compositional homogeneity. Raman results were analyzed by using Witec Project Plus Software. For better identification of vibrational modes, Fourier self-deconvolution (FSD) was carried out in the experimental average Raman spectra, following the same procedure as in [11, 12]. FSD is a mathematical tool that allows narrowing the vibrational modes, preserving their wavenumber and the integrated intensity of each line [40]. The parameters gamma and smoothing factor were fixed to 8 and 0.22, respectively.

X-ray absorption spectroscopy (XAS) measurements were carried out on the CSP and post-annealed samples to investigate the effect of the process and the organic solvent in the $SrFe_{12}O_{19}$ structure. Both X-ray absorption near-edge structure (XANES) and extended X-ray absorption fine structure (EXAFS) experiments were achieved at room temperature in transmission mode. Samples were characterized at the Fe and Sr K-edge at the BL22 CLÆSS beamline of the ALBA synchrotron facility in Cerdanyola del Vallès (Spain). The monochromator used in the experiments was a double Si crystal oriented in the (311) direction. Metal foils were measured and used as reference to calibrate the energy. $SrFe_{12}O_{19}$ starting powders were also measured as reference along with α-$Fe_2O_3$ and SrO powder references from transmitted photons. The phase composition of all samples was calculated from the linear combination fitting (LCF), from -20 to 200 eV with respect to each absorption edge, of the XANES spectra from standards. From EXAFS results, the Fourier transform (FT) was performed in the $k^3\chi(k)$ weighted EXAFS signal between 2.6 and 13.0 Å$^{-1}$ at the Fe K-edge and 3.0 and 11.5 Å$^{-1}$ at the Sr K-edge. EXAFS spectra were fitted in *R*-space in the range of 1.4–4.0 Å and 2.5–4.2 Å at the Fe and Sr K-edge, respectively, using the FEFFIT code [41,42]. For the fittings, the shift at the edge energy for each absorption edge was previously calculated from the starting $SrFe_{12}O_{19}$ powders and fixed for each sample. The coordination number *N*, the interatomic distance *R* and the Debye-Waller (DW) factors for each shell were considered as free parameters. For the fittings, a similar procedure to our previous work was followed [12]. At the Fe K-edge position, two shells were considered: a first one produced by the interaction of a Fe absorbing atom with six O atoms and a second one formed by two subshells of six Fe atoms each. At the Sr K-absorption edge, a three-shell model was considered: the first and second shell produced each by the interaction of a Sr absorbing atom with six O atoms and a third shell constituted by the interaction of a Sr absorbing atom with fifteen Fe atoms.

Magnetic properties of the sintered magnets were acquired by a physical property measurement system vibrating sample magnetometer (PPMS-VSM model 6000 controller, Quantum Design), under a maximum applied magnetic field of 5 T at room temperature.

## 3. Results and discussion

### 3.1. Morphological and structural characterization

The relative density values for the $SrFe_{12}O_{19}$ pieces sintered after the CSP and those post-annealed at 1100 °C for 2 h using glacial acetic acid, oleic acid and oleylamine as the CSP organic solvent are shown in Fig. 1B. Noticeable changes in the relative density of the samples after the CSP as a function of the pure organic solvent are found. The lowest relative densities about 72–74 % are obtained for OA and OL samples sintered with oleic acid and oleylamine, respectively, while the highest value, around 85 %, is acquired for the GAA sample prepared using glacial acetic acid, even at lower CSP temperatures. These findings indicate that the glacial acetic acid facilitates the mass transport and the dissolution-precipitation [32,43] during the CSP, while these processes are somewhat hampered using the oleic acid and the oleylamine with a low increase of the density value with respect to their green density value (around 60–65 %). Particularly, GAA sample presents sufficient mechanical integrity and can be considered as a dense piece, which is quite remarkable at this low temperature (190 °C), considering the high temperatures (>1200 °C) required in the traditional sintering processes [44,45]. Regarding the samples obtained after post-annealing process at 1100 °C for 2 h, relative densities increase to values around 92 % with respect to the theoretical density values, in all cases irrespective of the solvent employed. Therefore, under optimal sintering conditions, regardless of the solvent, sintered ceramics are obtained and can be employed in current magnetic applications.

By comparing the sintering process using pure organic solvents reported here with the standard 1 M aqueous acetic acid usually employed in the CSP [1] (see Fig. S2 in SI), it seems to inhibit the densification of $SrFe_{12}O_{19}$, yielding a relative density of 78 % after the CSP (aAA) and around 84 % after the post-annealing process (aAA1100), evidencing that a medium with a high acid character is necessary during CSP for densification, as Fig. 1B shows.

$SrFe_{12}O_{19}$ pieces processed at 1100 °C for 2 h using the same organic solvents and parameters without the CSP stage were also prepared, and named as GAA1100_CN, OA1100_CN and OL1100_CN using glacial acetic acid, oleic acid and oleylamine, respectively. In all cases, relative density values lower than 80 % (see Fig. 1B) with respect to the theoretical ones are obtained, highlighting the strong influence of the CSP in the densification during the sintering process of magnets.

The effect of organic solvent employed in the sintering route on the morphological characteristics of the ceramic pieces is investigated. Fig. 2 shows the FESEM images and the grain size distribution analysis for each sample after the CSP and the post-annealing process. In all cases, the grain size follows a log-normal distribution. As previously identified for $SrFe_{12}O_{19}$ particles sintered by CSP using glacial acetic acid (GAA), the solvent enables a local dissolution of the particle surface [12], identifying the coalescence of smaller grains in equiaxed grains and a large homogeneity of sample with a low open porosity (see Fig. 2). OA and OL samples prepared using oleic acid and oleylamine, respectively, show a skewed right distribution with a greater or lesser average particle size depending on solvent. The average grain size is quite similar for all solvents with values of 1.1 ± 0.4 μm, 0.8 ± 0.3 μm and 0.9 ± 0.4 μm for GAA, OA and OL, respectively.

After the post-annealing process of the CSP samples at 1100 °C for 2 h, the coalescence of smaller grains in equiaxed grains during the solid-state sintering is recognized. An average grain size around 1.0 ± 0.2 μm, 1.3 ± 0.4 μm and 1.1 ± 0.4 μm for GAA1100, OA1100 and OL1100, respectively, is found with platelet-shaped grains. The presence of straight grain boundaries indicates that the equilibrium in the sintering driving forces has been reached during the post-thermal treatment. A lower quantity of small grains is noted after the final sintering process, i. e. after the post-annealing step (see Fig. 2). The effect is quite similar employing oleic acid and oleylamine as the organic solvents in the sintering process, with more rounded and homogeneous grains for the GGA sample prepared with glacial acetic acid. It is important to remark that in the absence of the CSP step, the reported sintering process of dense hexaferrite magnets typically leads to average grain sizes larger than 3 μm [46].

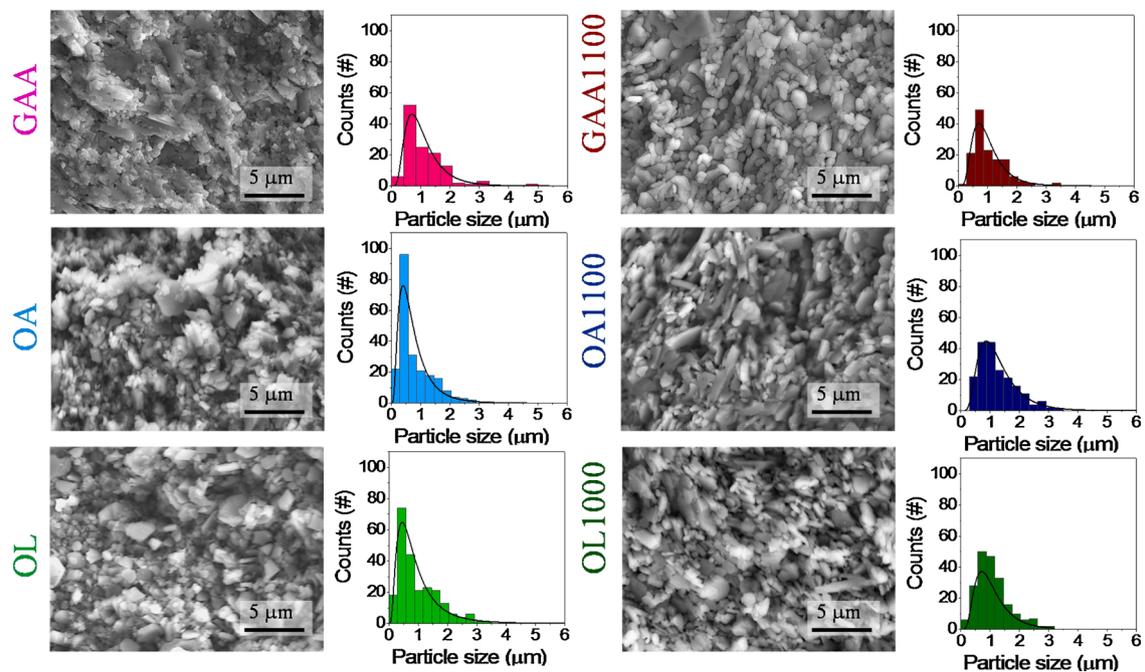

**Fig. 2.** FESEM micrographs and grain size distribution analysis of SrFe$_{12}$O$_{19}$ powders processed after the CSP using glacial acetic acid, oleic acid and oleylamine: GAA, OA and OL, and these annealed at 1100 °C for 2 h: GAA1100, OA1100 and OL1100. The grain size of the samples follows a log-normal distribution.

Refinements of XRD data allowed obtaining quantitative information related to the crystalline compositional phases presented in the samples depending on the solvent. Fig. 3A shows an example of the XRD patterns for the GGA and GAA1100 samples along with the Rietveld refinements of the XRD data. The refined weight fraction values for each identified phase in samples processed after CSP and post-annealed at 1100 °C are displayed in Fig. 3B. Specifically, for the GAA sample a partial decomposition of the ferrite phase is induced by the CSP obtaining 34 % of α-Fe$_2$O$_3$ with the rest as SrFe$_{12}$O$_{19}$. This transformation is recovered after the post-annealing process at 1100 °C identifying 100 % of SrFe$_{12}$O$_{19}$ for GAA1100 sample. For pieces prepared using oleic acid and oleylamine, no transformation of the ferrite phase is detected, with a 100 % of SrFe$_{12}$O$_{19}$ both after the CSP and the post-annealing process. Regarding the lattice dimensions and volume-averaged crystalline sizes, displayed in Fig. S3 of SI, no noticeable trends are identified.

Raman experiments were carried out to confirm the compositional crystalline phases obtained by XRD results and to identify other possible minority and/or amorphous phases. Fig. 4 shows the experimental average Raman spectra and the FSD spectra for ceramic magnets sintered after the CSP and after the post-annealing process. According to the group theory for the hexaferrites, 42 Raman vibrational modes are active: $11A_{1g} + 14E_{1g} + 17E_{2g}$ [47–49], of which some are identified in most of the sintered magnets related to the SrFe$_{12}$O$_{19}$ phase except for the ferrite processed after CSP using glacial acetic acid (GAA). In this last sample, we also identify vibrational bands related to α-Fe$_2$O$_3$ (see Fig. 4) [50–53], in which both the allowed phonon modes ($2A_{1g} + 5E_g$) and the activated LO $E_u$ mode associated with structural disorder in the α-Fe$_2$O$_3$ structure are exhibited. An assignment of the Raman bands is made taking into account previous works [47,51–53], and the recognized vibration modes are presented in Table S1 of SI. No Raman vibrational modes attributed to other phases are identified, corroborating the XRD data.

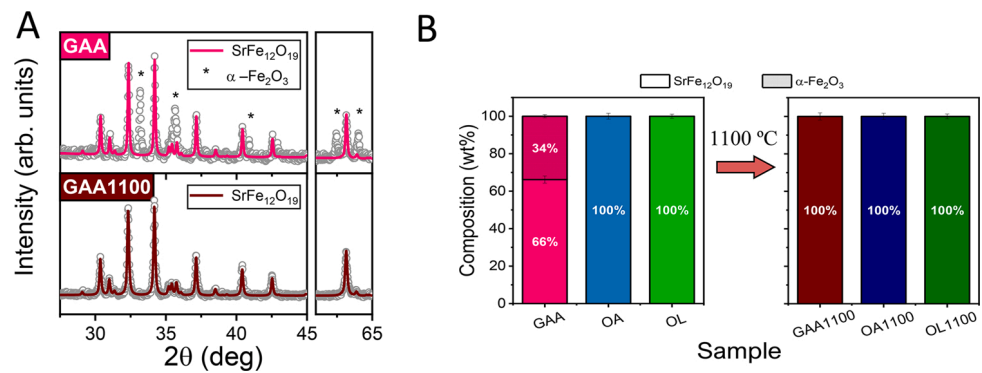

**Fig. 3.** A) Experimental XRD data (grey open circles) and Rietveld model built for the main phase in each case (solid line in colour) indicating the Bragg positions of the α-Fe$_2$O$_3$ phase (*). B) Phase composition in weight fraction extracted from Rietveld refinements of the XRD data measured for the sintered ceramics processed by the CSP varying the organic solvent (GAA, OA, OL) and by CSP followed by a post-annealing at 1100 °C for 2 h (GAA1100, OA1100, OL1100).

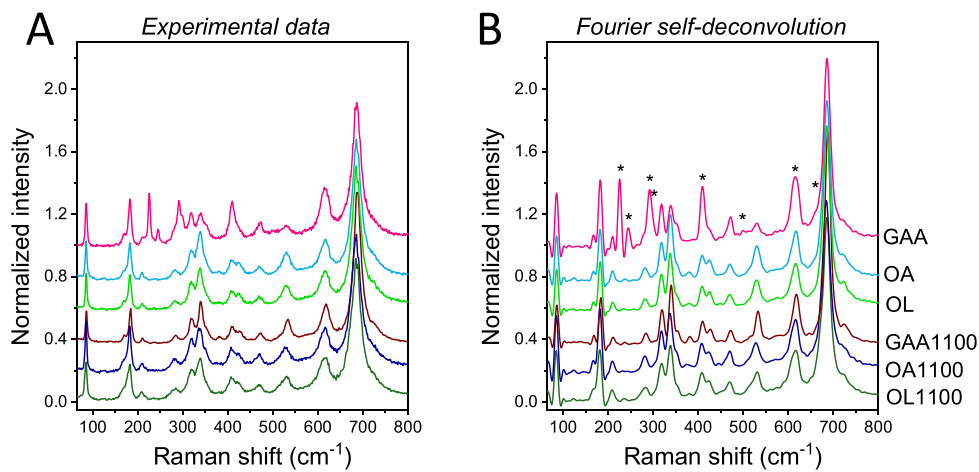

**Fig. 4.** A) Experimental average Raman and B) FSD spectra of the ferrite-based magnets sintered using different organic solvent after the CSP (GAA, OA, OL) and those plus a post-annealing stage at 1100 °C for 2 h (GAA1100, OA1100, OL1100). Spectra are presented from 65 to 800 cm$^{-1}$. Raman modes related to α-Fe$_2$O$_3$ phase are marked with an * on the Figure in the FSD spectrum. The rest of the vibrational bands corresponds to the Sr hexaferrite structure.

Variations in the relative intensity between the SrFe$_{12}$O$_{19}$ Raman bands are not significant, indicating low differences on average in the orientation of the hexaferrite platelets for the samples [47,49]. Specifically, for the GAA sample the changes in the intensity of Raman bands are associated with the proportion of α-Fe$_2$O$_3$ (see Fig. 4). In addition, variations in the position and full width high maximum (FWHM) of the Raman bands are noted, which can be attributed to several factors, including changes in the particle size or strain in the structure induced during the sintering process.

During the sintering process, in addition to the presence of minority phases not previously identified, important modifications on the short-range $SrFe_{12}O_{19}$ structure could have happened. This analysis was performed by XAS experiments at the Fe and Sr K-edge. Fig. 5A and B show the XANES signal at the Fe and the Sr K-absorption edge, respectively. In both edges, a similar XANES profile of the sintered samples to that of the $SrFe_{12}O_{19}$ starting powders is identified, except for GAA sample in which some resonances present certain modifications. By LCF of the XANES signal (see Fig. 5A and B), the compositional phases in each sample are determined, as shown Fig. 5C. After the CSP stage, the Sr ferrite is mostly retained using oleic acid and oleylamine, obtaining $SrFe_{12}O_{19}$ percentages around 91–95 % for all ceramic magnets, while glacial acetic acid partially destroys the ferrimagnetic phase, leaving a content of the $SrFe_{12}O_{19}$ phase of 46 %. Remarkably, along the $\alpha$-$Fe_2O_3$ phase already identified by XRD and Raman techniques in the GAA sample (around 26 % of $\alpha$-$Fe_2O_3$), here a fraction of SrO as secondary phase is obtained in all cases: around 28 % for GAA, 5 % for OA and 9 % for OL. This SrO phase may be considered as an amorphous compound produced during the CSP from the Sr cations located in the Sr ferrite structure. Usually, the presence of SrO is not identified in the investigations related to $SrFe_{12}O_{19}$-based permanent magnets. This study has made it possible its detection and what happens to the ferrite during the sintering process thanks to the sensitivity of XAS technique. The SrO phase was not probably identified by Raman technique due to its lower scattering cross-section with respect the $SrFe_{12}O_{19}$ and $\alpha$-$Fe_2O_3$ compounds. Other authors have also reported the low visibility of Raman modes for amorphous SrO [54]. As the sintering process is completed with the post-annealing stage, the ferrite phase is almost completely recovered in all cases with percentages >91 % and around 8 % of SrO, independently on the organic solvent employed in the process. An estimation of the crystalline phases is determined from XANES experiments considering the SrO phase as single amorphous phase, as shown in the SI (Table S2), corroborating the XRD quantified results.

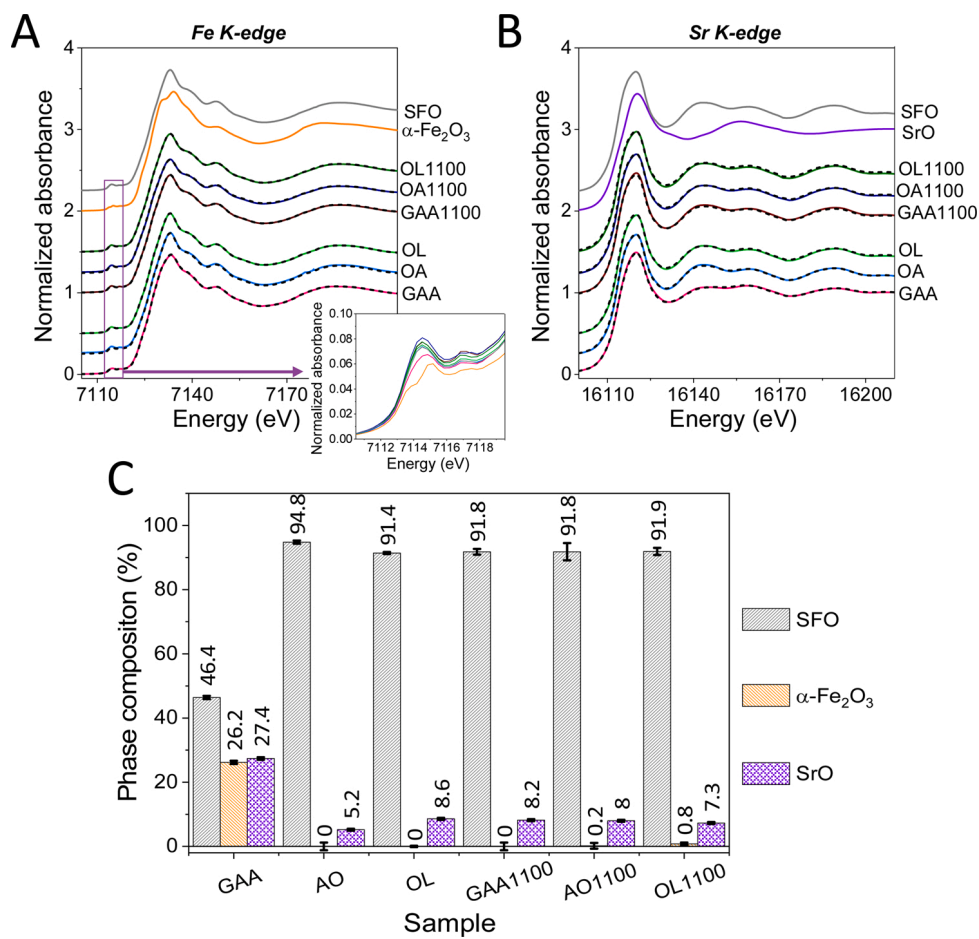

**Fig. 5.** XANES spectra (continuous lines) at the A) Fe and B) Sr K-edge and LCF (dashed lines) for all ceramic processed after the CSP (GAA, OA, OL) and the post-annealing process (GAA1100, OA1100, OL1100) using different pure organic solvent. C) Phase fraction obtained from the LCF for all ceramics sintered indicating the value on the Figure.

From the absorption edge positions in the XANES spectra, the average oxidation state of the absorbing Fe and Sr atoms is corroborated to be 3+ and 2+, respectively, calculated from the linear relationship between the absorption edge position (determined from the maximum of the first derivative curve) at the XANES regions at the Fe and Sr K-edge and the oxidation state of the absorbing atoms from standard references [12,55–57]. Specifically, in XANES region at the Fe K-edge, a pre-edge region is identified and attributed to a 1s → 3d transition [58], as inset depicted in Fig. 5A. No changes are observed at the pre-edge energy position of XANES spectrum indicating that the oxidation state of iron is the same for all samples [55], 3+, as stated above from the absorption edge position. However, modifications are observed at the pre-edge peak intensity around 7114.5 eV. The highest intensity is identified for the post-annealed samples, indicating a distortion of site symmetry for $Fe^{3+}$ cations respect to those of the starting $SrFe_{12}O_{19}$ powders that show a lower intensity with a coordination closer to the octahedron and less distorted [58]. The absorption pre-peak for the GAA sample exhibits an intermediate intensity between that of $SrFe_{12}O_{19}$ and $\alpha$-$Fe_2O_3$, corroborating the mix of both phases in the sample. Moreover, the pre-peak shape for the GAA sample becomes more asymmetric with respect to $SrFe_{12}O_{19}$ and its position slightly shifts by 0.2 eV towards higher energy, roughly in between the energy of the maximum pre-peak value of $SrFe_{12}O_{19}$ (7114.5 eV) and $\alpha$-$Fe_2O_3$ (7115.0 eV).

The short-range ordering of the Fe and Sr cations and the neighbor bond lengths were investigated by EXAFS and results for sintered magnets, as well as for $SrFe_{12}O_{19}$, SrO and $\alpha$-$Fe_2O_3$ references, are presented in Fig. 6 and Table S3 of SI. Similar results are obtained for all samples, except for GAA sample processed by CSP using glacial acetic acid. At the Fe K-edge, sample GAA shows structural parameters between those of the $SrFe_{12}O_{19}$ and $\alpha$-$Fe_2O_3$ structure: a clear reduction in the coordination and the distance of the third shell from $SrFe_{12}O_{19}$ reference towards the $\alpha$-$Fe_2O_3$ values is distinguished, corroborating the XRD, Raman and XANES results. However, for OA and OL samples, prepared using oleic acid and oleylamine, respectively, no modifications are identified. With respect to the Sr ions after the CSP, an intensity reduction at the three modeled coordination shells is identified for the GAA sample, while OA and OL samples show similar results to those of the $SrFe_{12}O_{19}$ reference (Table S3 in SI) with a slight decrease of the coordination number of the third shell (within the error).

As the post-annealing is performed, at the Fe K-absorption edge the ceramic magnets show similar EXAFS parameters between them and those obtained after the CSP, except for the GAA1100 sample that presents a shortening of the bond length distance at the third Fe-Fe shell. At the Sr K-edge, similar EXAFS results are also found (see Table S3 in SI).

With respect to the DW factor at both absorptions edges, no significant modifications are observed with respect to the $SrFe_{12}O_{19}$ reference powders, not with the CSP solvent or with the sintering step. At this stage, the reduction of the coordination number at both edges for the magnets sintered with respect to the starting $SrFe_{12}O_{19}$ powders (see Table S3 in SI) should be mentioned. If the variation is not related to the composition of the samples, it may indicate the loss of certain atoms into the structure inducing defects (e.g. vacancies) during the specific stage of the sintering process [12,32].

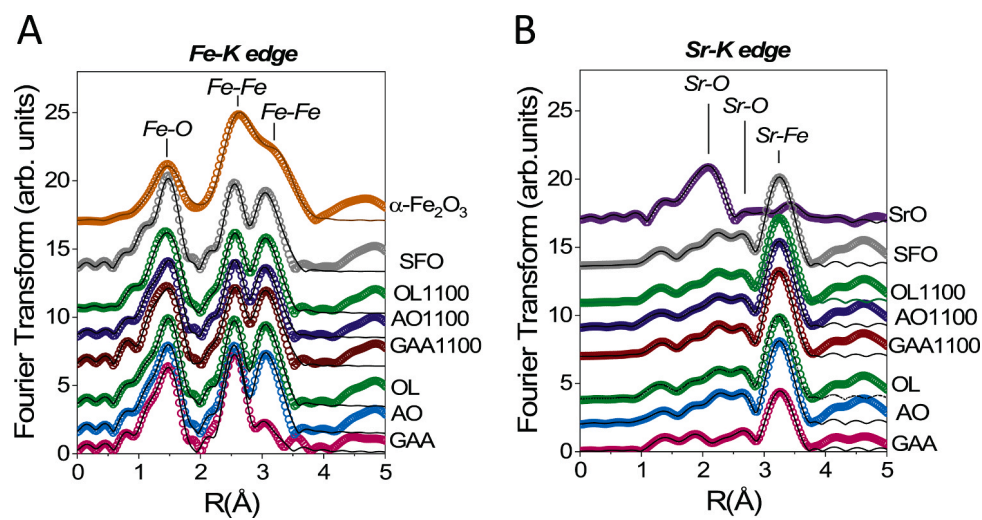

**Fig. 6.** Fourier transform modulus of the EXAFS spectra (dots) at the A) Fe and B) Sr K-edge and best-fitting simulations (continuous lines) of the magnets processed after the CSP (AAG, OA, OL) and after the post-annealing process (AGG1100, OA1100, OL1100) and the starting $SrFe_{12}O_{19}$ powders, along with $\alpha$-$Fe_2O_3$ and SrO references.

## 3.2. Magnetic properties

The magnetic properties of the ceramic magnets processed by CSP before and after the 1100 °C post-annealing are shown in Fig. 7. For all samples we observe a magnetic hysteresis signal with specific characteristics related to the morphological, compositional and structural properties that depend on the process stage and the organic solvent employed during the process.

After the CSP, the hysteresis loops present some differences depending on the organic solvent. The most important changes are induced in the GAA sample, with $H_C$ = 1.5 kOe and $M_S$ = 49.2 emu/g. The hard magnetic properties and the $M_S$ are reduced due to the partial transformation of the $SrFe_{12}O_{19}$ to the non-magnetic SrO phase and to $\alpha$-$Fe_2O_3$, iron oxide polymorph with a weak canted antiferromagnetic behavior at room temperature and a low saturation magnetization ($M_S \sim 1-2$ emu/g) [59]. For OA and OL samples, the magnetic response is very similar, with $H_C$ values around 2–2.2 kOe and $M_S$ at 70 emu/g. However, while the magnetic properties are attractive to be employed in certain applications, the mechanical integrity of the samples processed using oleic acid and oleylamine is not strong enough. As indicated above, post-annealing is required to restore the $SrFe_{12}O_{19}$ phase and attain greater density values. For these post-annealed samples, the magnetic properties are very similar irrespective of the organic solvent used during the process. All samples exhibit $M_S$ around 73 emu/g and $H_C$ larger than 2.1 kOe with the highest value of 2.8 kOe for the GAA1100 sample. This large $H_C$ value, compared to conventionally sintered samples, is the consequence of the smaller grain sizes, of the order of 1 μm, enabled by the sintering process at lower temperature employed here. Great efforts have previously been made to inhibit the grain growth during sintering of hexaferrites-magnets. For instance, Guzmán-Mínguez et al. have used $SiO_2$ as sintering additive during the process, achieving a maximum $H_C$ of 160 kA/m (2.0 kOe) [45]. The same authors have increased the $H_C$ up to 192 kA/m (2.4 kOe) by using a two-step sintering process in which the sintering time is reduced at the highest sintering temperature [46]. Jenuš et al. managed to reduce the sintering temperature down to 900 °C using spark plasma sintering technique but the $H_C$ of the resulting magnets did not surpass 2.1 kOe [60]. With respect to commercial ferrite magnets, several products with a wide range of coercivity and remanence values are available. Our best magnet (GAA1100 sample) reaches similar performance at lower sintering temperatures than the commercial ones from Hitachi metals NMF-6C series which offer $H_C$ = 2.8–3.3 kOe [61]. Therefore, it should be noted that: 1. the magnetic results of post-annealed samples sintered in this work are competitive with respect to commercial ferrite magnets without rare-earth-doping and upper to those sintered by conventional routes, 2. the sintering temperatures are reduced by around 8–12 % with respect to the conventional sintering routes in which temperatures >1200 °C are employed.

Another feature of these processed ferrite magnets to be highlighted is the degree of magnetic alignment exhibited between particles in the absence of an external magnetic field. A $M_R/M_S$ ratio ≥ 60 % is obtained for all samples except for GGA and OA samples (see Fig. 7), surpassing the predicted value for an assembly of randomly oriented, non-interacting particles [12,62]. This partial alignment is likely the consequence of the addition of the liquid organic solvents in combination with the applied pressure during the CSP, which reduces friction between the platelet-shaped ferrite particles and thus promotes their stacking and magnetic orientation.

## 4. Conclusions

Herein, the influence of the organic solvent employed during sintering has been evaluated in permanent magnets based on $SrFe_{12}O_{19}$ platelets, sintered by CSP plus a subsequent post-annealing at 1100 °C. Glacial acetic acid, oleic acid and oleylamine have been employed as organic solvents. Despite the partial transformation of the magnetic phase during the cold sintering process, which depends on the organic solvent, the final ceramic magnets present > 90 % of the ferrimagnetic phase with a low fraction of crystalline $\alpha$-$Fe_2O_3$ and amorphous SrO. Irrespective of the solvent, after the sintering process, permanent magnets with relative density values around 92 % of the theoretical density and competitive $H_C$ values are obtained. These findings thus lead to interesting magnetic properties as permanent ferrite magnets. Moreover, the reduction of the sintering temperature enabled by the CSP not only yields smaller grain sizes, but opens the door as well to the consolidation of multi-phase ferrite-based magnets, such as hard-soft composites, that are otherwise altered at higher temperatures.

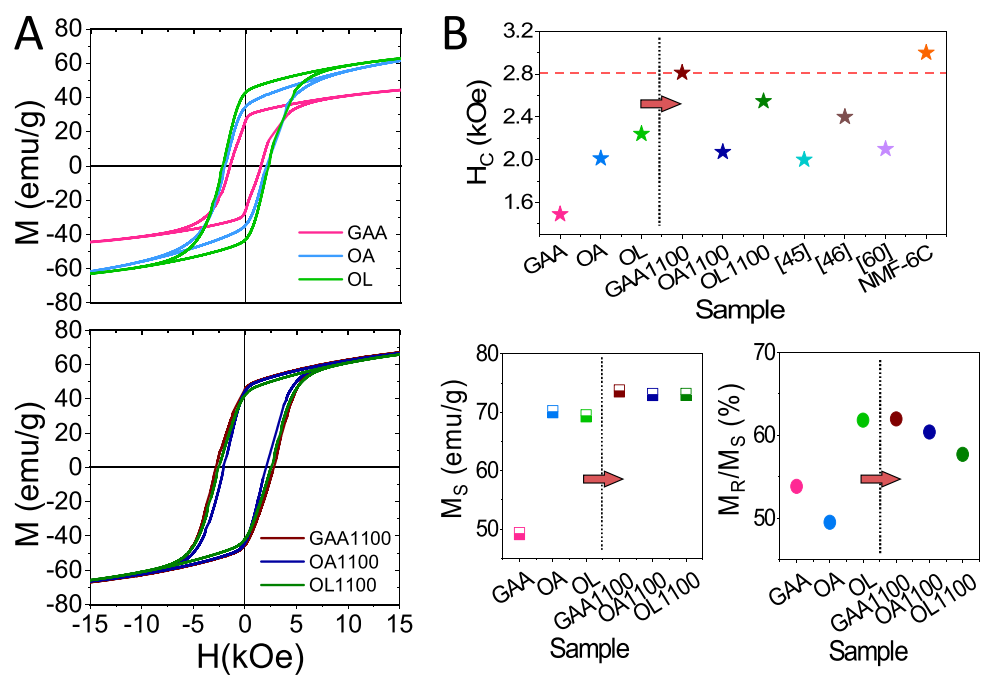

**Fig. 7.** A) Hysteresis loops representing the magnetic response of ceramic magnets processed by CSP using glacial acetic acid, oleic acid and oleylamine: GAA, OA and OL, and these post-annealed at 1100 °C for 2 h: GAA1100, OA1100 and OL1100. B) $H_C$, $M_S$ and $M_R/M_S$ ratio derived from the hysteresis loops. Coercivities are compared with those values reported by Guzmán-Mínguez et al. [45,46], Jenuš et al. [60] and the commercial magnet from Hitachi metals NMF-6C series [61].

**Author contributions**

All authors have given approval to the final version of the manuscript.

**Declaration of Competing Interest**

The authors report no declarations of interest.

**Acknowledgments**

This work has been supported by the Ministerio Español de Ciencia e Innovación (MICINN) through the projects MAT2017-86540-C4-1-R and RTI2018-095303-A-C52, and by the European Commission through Project H2020 No. 720853 (Amphibian). Part of these experiments was performed at the CLAESS beamline at ALBA Synchrotron with the collaboration of ALBA staff. A.S. acknowledges financial support from Comunidad de Madrid for an "Atracción de Talento Investigador" Contract (2017-t2/IND5395). C.G.-M. and A.Q. acknowledge financial support from MICINN through the "Juan de la Cierva" Program (FJC2018-035532-I) and the MICINN through the "Ramón y Cajal" Contract (RYC-2017-23320), respectively.

**Appendix A. Supplementary data**

Supplementary material related to this article can be found, in the online version, at doi:https://doi.org/10.1016/j.jeurceramsoc.2021.10.062.